\documentclass[preprint]{elsarticle}
\usepackage{graphicx}
\usepackage{xspace}
\usepackage{alltt}
\usepackage{url}            
\usepackage{xspace}
\usepackage{bchart}
\usepackage{setspace}
\newboolean{showcomments}
\setboolean{showcomments}{true}
\ifthenelse{\boolean{showcomments}}
  {\newcommand{\bnote}[2]{
	{\color{cyan}\fbox{\bfseries\sffamily\scriptsize#1}
    {\sf\small$>$\textit{#2}$<$}}
   }
   
  }
  {\newcommand{\bnote}[2]{}
   
  } 

\newcommand{\etal}{\emph{et al.}\xspace}

\newcommand\tk[1]{\bnote{Tobias}{#1}}

\newcommand{\sepp}{$>>$\xspace}

\newcommand{\stBar}{$\mid$}
\newcommand{\ret}{\^{}}
\newcommand{\myparagraph}[1]{\noindent\textbf{#1.}}
\newcommand{\eg}{e.g.,\xspace}
\newcommand{\ie}{i.e.,\xspace}
\newcommand{\co}[1]{{\small\textsf{#1}\xspace}}
\newcommand{\secref}[1]{Section \ref{sec:#1}}
\newcommand{\seclabel}[1]{\label{sec:#1}}

\newenvironment{tcode}
    {\begin{quote}\footnotesize\sf\begin{tabbing} xx\=xxxxxx\=xxxxxx\=xxxxxx\=xxxxxx\=xxxxxx\=xxxxxx\=xxxxxx\=xxxxxx\=xxxxxx\=xxxxxx\=xxxxxx\=xxxxxx\=\kill}
    {\end{tabbing}\end{quote}}

\newcommand{\figref}[1]{Figure~\ref{fig:#1}}
\newcommand{\figlabel}[1]{\label{fig:#1}}
\newcommand{\tablabel}[1]{\label{tab:#1}}

\usepackage[normalem]{ulem} % for \sout
\usepackage{xcolor}
\usepackage{color}

\newcommand{\ace}[1]{{\emph{#1}}}

\newenvironment{aceenv}
    {\begin{quote}\it\small\raggedright\setlength{\leftskip}{0cm}\parindent=-0.3cm\hspace{\parindent}\parskip=0pt plus 1pt}
    {\end{quote}}

\begin{document}

%SPRINGER
%\title{Verifiable Code Documentation in Controlled Natural Language}
%\author{Tobias Kuhn \and Alexandre Bergel}
%\authorrunning{T. Kuhn \and A. Bergel}
%\institute{%University of Chile, Santiago, Chile\\
%\co{\href{http://www.tkuhn.ch}{www.tkuhn.ch} / \href{http://www.bergel.eu}{www.bergel.eu}}\\
%}

%ACM
%\conferenceinfo{ESEC/FSE}{2011}

%\title{Verifiable Software Documentation in\\Controlled Natural Language}
\title{Verifiable Source Code Documentation in\\Controlled Natural Language}

%\numberofauthors{2}
%\author{
%
%Tobias Kuhn   ~~~~~   Alexandre Bergel\\
%University of Chile, Santiago, Chile\\ [1 ex]
%\href{http://www.tkuhn.ch}{http://www.tkuhn.ch}~~~~~\href{mailto://kuhntobias@gmail.com}{kuhntobias@gmail.com}\\
%\href{http://bergel.eu}{http://bergel.eu}~~~~\href{mailto://abergel@dcc.uchile.cl}{abergel@dcc.uchile.cl}\\
%\href{mailtto:alexandre.bergel@inria.fr}{alexandre.bergel@inria.fr}
%        \email{{\small \href{mailto:alexandre.bergel@inria.fr}{alexandre.bergel@inria.fr}}}
%        \email{\url{http://www.bergel.eu}}
%
%\alignauthor
%  Tobias Kuhn\medskip\\
%  \affaddr{Department of Intelligent Computer Systems}\\
%  \affaddr{University of Malta, Msida, Malta}\smallskip\\
%  \href{http://www.tkuhn.ch}{http://www.tkuhn.ch}\\
%  \href{mailto://kuhntobias@gmail.com}{kuhntobias@gmail.com}
%\alignauthor
%  Alexandre Bergel\medskip\\
%  \affaddr{Computer Science Department}\\
%  \affaddr{University of Chile, Santiago, Chile}\smallskip\\
%  \href{http://bergel.eu}{http://bergel.eu}\\
%  \href{mailto://abergel@dcc.uchile.cl}{abergel@dcc.uchile.cl}
%}

\author{Tobias~Kuhn$^1$ and~Alexandre~Bergel$^2$\\\vspace{0.2cm}
$^1$ Chair of Sociology, in particular of Modeling and Simulation, ETH Zurich\\ \href{http://www.tkuhn.ch}{http://www.tkuhn.ch} ~~~ \href{mailto://kuhntobias@gmail.com}{kuhntobias@gmail.com} \\\vspace{0.2cm}
$^2$ Pleiad Lab, Department of Computer Science (DCC),
University of Chile \href{http://bergel.eu}{http://bergel.eu}}

\date{}

%\author{
%\IEEEauthorblockN{Tobias Kuhn}
%\IEEEauthorblockA{Department of Intelligent Computer Systems\\
%University of Malta\\
%Msida, Malta\\
%\href{http://www.tkuhn.ch}{http://www.tkuhn.ch} / \href{mailto://kuhntobias@gmail.com}{kuhntobias@gmail.com}}
%\and
%\IEEEauthorblockN{Alexandre Bergel}
%\IEEEauthorblockA{Computer Science Department\\
%University of Chile\\
%Santiago, Chile\\
%\href{http://bergel.eu}{http://bergel.eu} / \href{mailto://abergel@dcc.uchile.cl}{abergel@dcc.uchile.cl}}
%}

\newcommand{\spp}{~~~~~~~}

\begin{frontmatter}

%\IEEEcompsoctitleabstractindextext{%  

\begin{abstract}
%Problem
Writing documentation about software internals is rarely considered a rewarding activity. It is highly time-consum\-ing and the resulting documentation is fragile when the software is continuously evolving in a multi-developer setting. Unfortunately, traditional programming environments poorly support the writing and maintenance of documentation.
%Why the problem is a problem
Consequences are severe as the lack of documentation on software structure negatively impacts the overall quality of the software product. 
%What is the surprising idea
We show that using a controlled natural language with a reasoner and a query engine is a viable technique for verifying the consistency and accuracy of documentation and source code. Using ACE, a state-of-the-art controlled natural language, we present positive results on the comprehensibility and the general feasibility of creating and verifying documentation.
%What are the consequences
As a case study, we used automatic documentation verification to identify and fix severe flaws in the architecture of a non-trivial piece of software.
Moreover, a user experiment shows that our language is faster and easier to learn and understand than other formal languages for software documentation.
\end{abstract}

%\begin{IEEEkeywords}
%software documentation; semantic wiki; controlled natural language;
%\end{IEEEkeywords}}

%\IEEEdisplaynotcompsoctitleabstractindextext
%\IEEEpeerreviewmaketitle

%\category{D.3.3}{Programming Languages}{Language Constructs and Features}
%\category{D.1.5}{Programming Languages}{Object-oriented Programming}
%\terms{Language, Design}
%\keywords{Multi-language system, interoperability, SmalltalkLite, JavaLite, dynamic languages}

\end{frontmatter}

%:%%%%%%%%%%%%%%%%%%%%%%%%%%%%%%%%%%%%%%%%%%%%%%%%%%

\section{Introduction} \seclabel{introduction}

%Context
Software documentation is commonly understood as any description of software attributes, and documenting software is a critical factor to success~\cite{Paul03a}. Glancing at the website of the ten most popular SourceForge software programs\footnote{VLC, eMule, Vuze, Ares Galaxy, Smart package, 7-Zip, Notepad++ Plugin Manager, PortableApps.com, FileZilla, and MinGW are the most downloaded software at the time this article is being written. See \url{http://sourceforge.net/}} easily demonstrates that: all software comes with a fair amount of documentation, including screenshots, tutorials, and FAQs.

%Problem
However, scrutinizing the documentation of these popular programs reveals that the offered documentation, with extensive installation procedures and problem troubleshooting instructions, is essentially focused on end users. None of the projects we studied present a clear description of their internal structure. The source code of eMule, for example, weighs more than 8 MB with 260\,000 lines of code. The extensive website of eMule contains many documentation files\footnote{\url{http://www.emule-project.net/home/perl/help.cgi?l=1}}, but none of them are targeted toward developers. 

This situation is not an exception, especially when the documenting of the software internals is not a strong requirement~\cite{Jako05a}, as is often the case with open source software. There are good reasons for this, including the lack of resources, the fragility of the documentation~\cite{Marc03b,Pier02b}, and the lack of immediate benefits. Keeping the internal documentation up to date with the source code is particularly difficult~\cite{Shi11a}. %Moreover, laking documentation about software internal is probably exacerbated in presence of agile technique. Unit tests are widely recognized by agile communities as the main source of code documentation~\cite{}. 
This article proposes an effective and convenient solution to write documentation that is interpreted as a formal model, is automatically checked, and is kept in sync with the source code.

%Solution
We propose the use of controlled natural language \cite{wyner2009cnlmain,kuhn2013cl} to represent software structure and its relation to the outside world, which leads to documentation that is automatically verifiable. The idea is to let developers write statements in a subset of English, describing the design and the environment of software that is otherwise nowhere explicitly expressed in the source code (\eg \emph{Simon is a developer}, \emph{Shapes is a component}, \emph{Every subclass of MOShape belongs to Shapes}, and \emph{Simon maintains every class that belongs to Shapes}).

Among the different kinds of controlled natural languages \cite{kuhn2013cl}, our approach relies on the type of languages that provide a natural and intuitive representation for formal notations. The statements of such languages look like natural language (such as English), but are controlled on the lexical, syntactic, and semantic levels to enable automatic and unambiguous translation into formal logic notations. We propose to use such natural-looking statements to write documentation statements about the source code structure. Ideally, these documentation statements should be next to the source code, e.g. in code comments, and written, read, and evaluated in the IDE of the developer. In this way, changes in the source code or its documentation that would introduce an inconsistency can be automatically detected. Furthermore, the logical structure of the documentation is accessible to query engines, via questions written in the same controlled natural language.

%Results, Contribution
We evaluate our approach using the language Attempto Controlled English (ACE) \cite{fuchs2008reasoningweb,decoi2009rewerse} and the semantic wiki engine AceWiki \cite{kuhn2008swui}. To test the feasibility of our approach, we apply it to document Mondrian, an open-source visualization engine, and to improve its design. In addition, we conduct a user experiment to test how controlled natural language compares to other formal languages for source code documentation and querying.

In this article, we investigate the following two research questions:
\begin{itemize}
\item[A --] \emph{Can controlled English be efficiently used to document software systems?}
\item[B --] \emph{Is controlled English easier to understand than other formal languages for software documentation?}
\end{itemize}

%:HERE
A growing amount of empirical evidence~\cite{Sill06a} shows that programmers spend a significant amount of time understanding source code and navigate between source code of software components. The generality of available software engineering tools and their poor performance are often pointed out as the culprit for engineering tasks being carried out in a suboptimal fashion~\cite{Stor97a,DeAl08a}. Informal, broken, or poorly conceived documentation seem to be at the root of many difficulties encountered in software engineering tasks~\cite{Sill06a,Rigb13a,Zhon13a}.
The two research questions we have stated above explore a new direction: documentation should be written in restricted natural language and automatically verified against the actual source code. %interaction between the programmer and the programming environment for source code querying and navigation.

We tackle the first research question by documenting a significant amount of source code in ACE using AceWiki. The second research question is investigated by conducting a user experiment with 20 participants.

%We empirically demonstrate that documentation written in ACE has many benefits over current approaches.

%Documentation written in a controlled natural language like ACE constitutes a formal model of the source code and its environment. Using a reasoner, the documentation can be checked for consistency, invariants can be verified, queries can be automatically answered, and different kind of data can be automatically extracted and used by other tools.

%The main innovations and contributions of this paper are summarized as follows:
%\begin{itemize}
%\item representation of the source code structure in controlled English
%\item source code annotation and documentation in controlled English
%\item querying the source code and its documentation in controlled English 
%%\item as a consequence, source code structure, queries, annotations, and documentation are represented in the same language
%\item prototypical implementation for automatic synchronization of documentation with the application source code
%\end{itemize}

We would like to stress that this article is \emph{not} about providing yet another solution for querying source code. Effective solutions have been proposed already, using logic programming~\cite{Wuyt01a} or natural language~\cite{wuersch2010icse}. Reverse-engineering and reengineering environments typically offer sophisticated way to query and extract information from source code. 
Our approach combines source code modeling techniques with general logic expressions for documentation, all accessed via a control natural language and accessible to formal reasoning. ACE has been successfully applied in areas such as the Semantic Web, however its application to software documentation is new and has not been considered as far as we are aware of.
%As far as we are aware, documenting software (even partially) in a controlled natural language has not been considered what-so-ever by academia or industry. 

We also would like to point out that our approach focuses on \emph{internal} code documentation.
Software documentation can describe a wide variety of aspects, including system requirements, architecture, design decisions, and implementation details. In this article, we focus on the low level source code structure where documentation statements talk about source code elements such as classes and methods, but express things that are not visible in the source code. This includes connections to the outside world, such as who is responsible for which classes and what external devices a particular piece of code depends on.

In order to answer the research questions introduced above, this article makes the following contributions:
\begin{enumerate}[i --]
\item it presents a concrete approach to document source code with a controlled natural language;
\item it shows how source code modeling can be combined with controlled natural language and automated reasoning; 
\item it evaluates the approach on a case study of documenting a medium-sized application;
\item it presents the results of a user experiment for the comparison of the understandability of different languages for source code documentation.
\end{enumerate}
These contributions consist of applying a state-of-the-art approach on natural language processing to address the recurrent  software engineering problem of writing understandable and verifiable software documentation.

This article is organized as follows: first, we motivate our approach and provide some background information (\secref{motivbackgr}), before introducing the specifics of our approach (\secref{approach}); then, we describe our implementation that is based on AceWiki (\secref{implementation}); subsequently, the case study on Mondrian is detailed (\secref{casestudy}), the design of the user experiment and its results are explained (\secref{experiment}), a number of points are discussed (\secref{discussion}), before our work is put in a broader perspective (\secref{relatedwork}) and we draw some conclusions from this work (\secref{conclusion}).

%\paragraph{Lake of documentation} Surveying the most 10 popular applications in sourceforge reveals that little documentation about the application design and other internal is provided. Although nothing implies that a popular application needs to have its design described, popular applications are large and involve many developers. To follow with the example introduced earlier, 19 people are working on developing, testing and debugging the 260 000 lines of code of eMule. 

%:%%%%%%%%%%%%%%%%%%%%%%%%%%%%%%%%%%%%%%%%%%%%%%%%%%
\section{Motivation and Background} \seclabel{motivbackgr}

Even though the benefits of code documentation are well known~\cite{Royc70a,Stet11a}, many popular applications have poor or outdated documentation. Donald Knuth has for long advocated a radical change toward the way we build software programs~\cite{Knut92a}:
\begin{quote}
Let us change our traditional attitude to the construction of programs: Instead of imagining that our main task is to instruct a computer what to do, let us concentrate rather on explaining to human beings what we want a computer to do.
\end{quote}
Knuth argues that a software program should be conceived as a work of literature. Unfortunately, software engineers are far from being that disciplined. Several studies show that internal documentation is considered important but is rarely written~\cite{Stet11a,Souz05a}.

The need of better documentation is often perceived as one of the principal reasons for difficulties encountered in software development~\cite{Sill06a,Zhon13a,Buse10a,badger2006eosm}.
Empirical studies~\cite{Sill06a} show that an average programmer spends a significant amount of time ``finding initial points in the code that were relevant to the task'', finding ``more information relevant to the task'', and understanding graphs of interacting code. 

Recent work in software documentation~\cite{Rigb13a,Zhon13a} has focused on relating source code to informal descriptions (\eg API descriptions or StackOverflow posts). In this way, obsolete or erroneous pieces of documentation can be identified, and social networks can be analyzed to look for discussions associated to a particular part of the source code.
Current approaches based on informal documentation mostly use simple text pattern matching, \eg identifying typos in class names~\cite{Zhon13a} and dividing camel-cased terms into compound terms~\cite{Bacc10a}. In contrast, our two research questions we stated earlier (\secref{introduction}) focus on the use of a controlled natural language for software documentation, which allows us to carry out sophisticated analyses (in particular for queries and checks) that are not possible in a reliable way with uncontrolled natural language.

We identified three main challenges to overcome in order for a documentation system based on a control natural language to be practical, effective and reliable:\\

%\noindent\textbf{Use of a natural language.} A documentation is best written in a language that is 

\noindent\textbf{Validation.}
Documentation is only useful if it correctly reflects the implementation, otherwise it can be counterproductive. Documentation therefore needs to be regularly validated.\\

\noindent\textbf{Immediate feedback.}
The documentation of source code is usually fragile with respect to code changes. A small change can invalidate a considerable part of the documentation. Only with immediate feedback, divergences between documentation and source code can be quickly identified, which is essential to prevent desynchronization.\\

\noindent\textbf{Minimal impact on software engineers.}
To motivate software developers to write sufficient and good documentation, the effort for writing documentation should be minimal and should not disrupt their work flow.\\

Numerous approaches have been proposed to address these challenges. Their impact, however, is rather limited so far.

Interface Definition Languages, for instance, are specification languages to describe software components (\eg WebIDL\footnote{\url{http://www.w3.org/TR/WebIDL}}). Their application range is commonly limited to the very technical corner, such as remote procedure invocations. A description is meant to be read by a compiler, not by a human being. IDLs can be easily verified and can produce immediate feedback, but the extra work for software engineers is considerable. To document software in IDL, one has to first learn the IDL syntax, become familiar with the IDL tools, and understand how to detect and interpret a mismatch between documentation and source code. 

Modern programming environments offer a large set of tools to navigate and query source code. Such queries are usually expressed in a proprietary domain specific language\footnote{\eg \url{http://browsebyquery.sourceforge.net}} or in a kind of SQL-like language.\footnote{\eg \url{http://www.eclipse.org/modeling/emf/?project=query}} Punctual queries on the source code can be used to verify structural properties, but they cannot easily be exported and stored for documentation purposes.

A fair amount of previous work exists on how to reduce the documentation effort that has to be provided by software engineers. For example, Sridhara \etal~\cite{Srid11a} generate comments for method arguments, and Buse \etal~\cite{Buse08a} automatically document exceptional events in natural language.
%In either case, around 85\% of the generated comments are evaluated as accurate.
These approaches only cover a rather narrow range of possible documenation statements. One can argue that the crucial role of documentation is to point out things that cannot be seen in the source code, and the above approaches are no help in this respect.
%However, these techniques use the source code as an oracle, which means that the accuracy of the source code is never questioned. The source code is assumed to be correct and documentation is automatically generated out of it. As we will illustrate later, this premise does often not hold because software source code is easily deteriorated.

%It has also been shown that succinct human-readable documentation are synthesizable from a software change~\cite{Buse10a}. However, it is unclear whether such technique is applicable other than helping write version log commit.

To deal with the challenges outlined above, we propose to use controlled natural language. Controlled natural languages are restricted versions of natural languages (\eg English) that are easily processable by computers. Grammar and dictionaries are restricted in such a way that ambiguities are reduced or (as in our case) completely eliminated. Recently, controlled natural languages have started to be employed to query software source code~\cite{wuersch2010icse}. 

In the subsequent sections, we present our approach on applying controlled natural languages to software documentation. With the help of a source code model, a controlled natural language, a proper editing environment, and a reasoning engine, software engineers can write documentation that can be validated, leads to immediate feedback, and imposes only minimal learning efforts on the software engineers.

%:%%%%%%%%%%%%%%%%%%%%%%%%%%%%%%%%%%%%%%%%%%%%%%%%%%
\section{General Approach} \seclabel{approach}

In what follows, we first present the general workflow of our approach (\secref{workflow}). Then, we describe how software source code is represented (\secref{sourcecode}), before we explain the main operations available to developers who have to document software while ensuring the produced documentation is synchronized with the source code (Sections \ref{sec:verifiabledocumentation} -- \ref{sec:consistency}).

%:==============

\subsection{Workflow}\seclabel{workflow}

\begin{figure}[!]
\begin{center}
\includegraphics[scale=0.4]{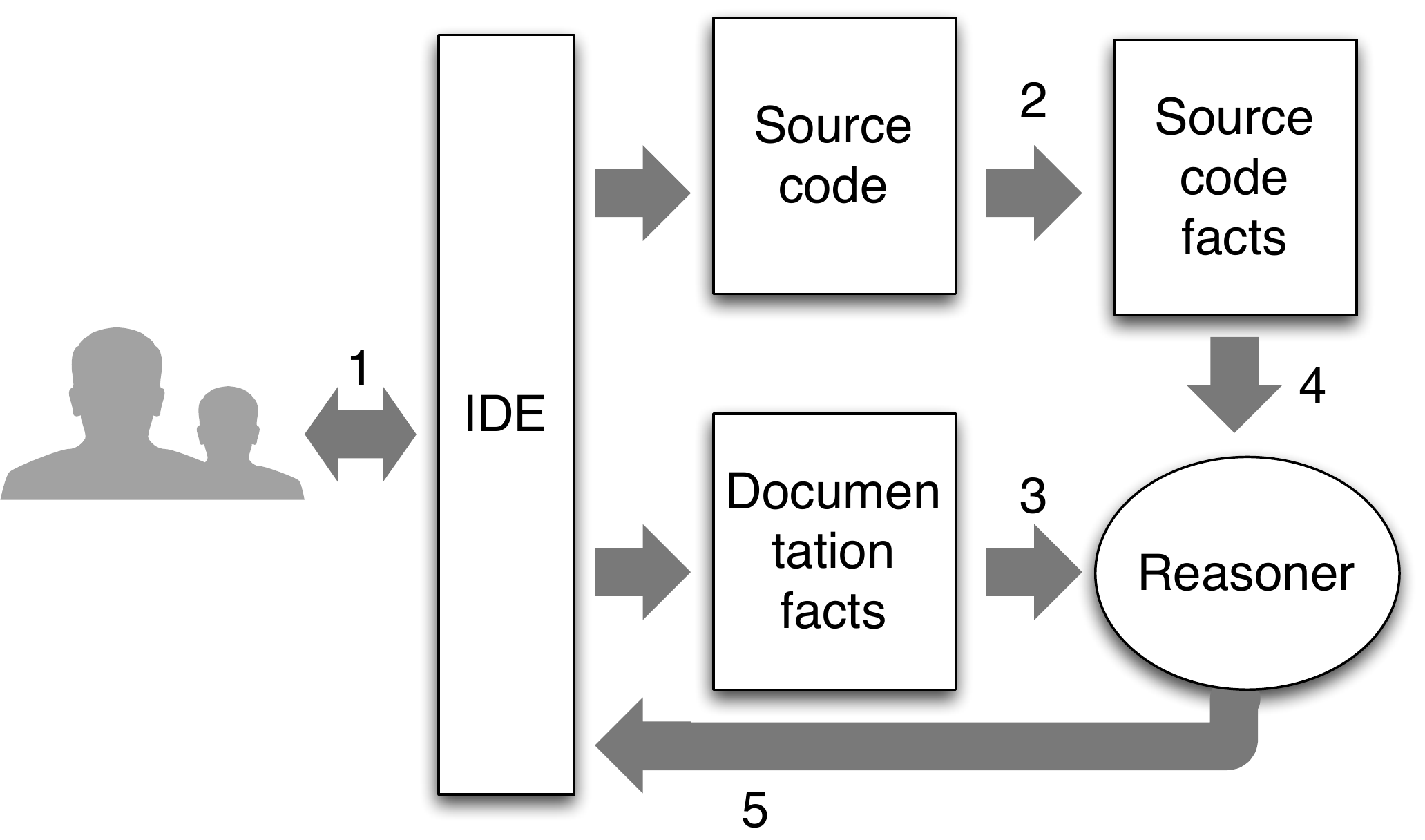}
\caption{Workflow of our approach}
%\caption{Ideal workflow of our approach. Currently, we do not have a single IDE as shown in the diagram, but use a semantic wiki alongside a classical IDE. This is, however, just a temporary shortcoming and not a design decision of our approach.}
\figlabel{workflow}
\end{center}
\end{figure}

The different steps of our approach are shown in \figref{workflow}:

\begin{itemize}
\item \emph{Step 1}: Developers write documentation statements in controlled natural language while writing the source code, ideally using the same integrated development environment. Such documentation statements can, for example, be written in specifically tagged source code comments, which makes them easily extractable.
%The IDE processes the source code and the documentation, offering syntax highlighting, code completion, and spellchecking.

\item \emph{Step 2}: Facts about the code structure are extracted from the source code. This is based on a source code model that is specific to the given programming language. Section \secref{sourcecode} explains the details of such source code models.
%The source code reification is done by the Moose\footnote{\url{www.moosetechnology.org}} tool suite. 

\item \emph{Steps 3 and 4}: The manually written documentation statements and the automatically generated source code statements are given to a logic reasoner.
% The documentation and the source are passed to the logic reasoner in term of controlled natural language statements.

\item \emph{Step 5}: The reasoner performs inference tasks like consistency checking and query answering, and present the results to the developer via the IDE. This is explained in the sections \ref{sec:queries} and \ref{sec:consistency}.
\end{itemize}

%:==============

\subsection{Source Code Model}\seclabel{sourcecode}

In order to link the documentation to the source code, a source code model has to be defined. In our approach, this model is automatically extracted from the source code, and can later be used as a basis to manually write more general documentation statements. The generation of the source code model is just a prerequisite for the second part of our approach, but is not particularly interesting or novel on its own.

Examples of core concepts of such a source code model could be \ace{code element}, \ace{class}, \ace{method}, \ace{package}, or \ace{interface}. Core relations could be \ace{direct subclass of}, \ace{defines}, \ace{invokes}, or \ace{instantiates}. We will use the language ACE for the examples to be shown below, but any other controlled natural language of sufficient expressiveness could be used. In any case, these concepts and relations should be defined together with morphological variants like \ace{define} and \ace{defined}, in order to be able to produce well-formed sentences. Restrictions on these core concepts and relations could be defined as follows:

\begin{aceenv}
Every class is a code element.\\
Every method is a code element.\\
No class is a method.\\
If X defines Y then X is a class.\\
If X defines Y then Y is a method.\\
\end{aceenv}

When a word like \ace{class} is just introduced, the reasoner does not know anything about the concept of a \ace{class}. By adding statements like the ones above, the word \ace{class} is formally linked to other words (by means of first-order logic), which captures (part of) the meaning of the word and enables the automatic calculation of inferences. The first example states that \ace{class} is a subconcept of \ace{code element}; the third one states that it does not overlap with the concept \ace{method}; and the fourth connects it to the relation \ace{defines}, saying that only classes can define something.
%\ab{I like this paragraph. However, what is the internal structure of ACE? Maybe you can simply put a bib ref after ``is formally linked to other words''}
In this way, we tell the reasoner about the domain of object-oriented source code (``we'' means us as system developers, not end users).
% Without that, the reasoner would not know what a class or a method is and would not be able to deduce any meaningful inferences.

Based on this model, we can automatically extract source code statements, for example:
\begin{aceenv}
EventHandler is a class.\\
EventHandler is a direct subclass of Handler.\\
EmergencyHandler is a class.\\
EmergencyHandler is a direct subclass of EventHandler.\\
EmergencyHandler-isActive is a method.\\
EmergencyHandler defines EmergencyHandler-isActive.\\
\end{aceenv}

Again, it is important to note that users are not supposed to write such statements, but they are automatically extracted and ideally synchronized in real time as the source code changes. As we will see, these statements together with the source code model serve as a scaffold for user-defined documentation statements.

%The structure of Mondrian is described using 2 050 facts (148 facts for class declarations and 1 902 facts for method declarations).
In addition to the statements shown above, we can extract statements about the invocation that occurs between methods:
\begin{aceenv}
EmergencyHandler-trigger invokes AlarmDesk-setAlarm.\\
AlarmDesk-setAlarm invokes Broadcaster-sendMessage.\\
\end{aceenv}
%
%For archive:
%| str |
%str := FileStream fileNamed: 'AceSyntax.txt'.
%self asArray do: [:cls | 
%	str nextPutAll: cls name , ',', cls superclass name; cr.
%	cls methods do: [:m |
%		str nextPutAll: cls name, ' defines ', cls name, '-', m name; cr.
%		m invokedMethods do: [:m2 |
%			str nextPutAll: cls name, '-', m name, ' invokes ', m2 parentType name, '-', m2 name; cr ].
%	] ].
%str close 
%
Based on these core concepts and relations, we can define additional relations for convenience reasons, like \ace{subclass of}, \ace{direct superclass of}, \ace{superclass of}, \ace{method of}, \ace{class of}, and \ace{uses}:

\begin{aceenv}
If X defines Y then Y is a method of X.\\
If X is defined by something that belongs to Y then X is a method of Y.\\
If X invokes something that is defined by Y then X uses Y.\\
If X defines something that uses Y then X uses Y.\\
\end{aceenv}

Altogether, this shows how the general structure of source code can be formally represented in controlled natural language, which is the basis for expressing actual documentation statements.
%As mentioned above, this is an important step of our approach because it allows us to link documentation statements to the source code. In this way, automatic reasoning tasks can be performed on this formal model.

%:==============

\subsection{Verifying Documentation Written in ACE}\seclabel{verifiabledocumentation}

We now come to the important and interesting part of our approach: user-defined documentation. The described source code model is the starting point, which can be extended by additional ontological entities (i.e. introducing new words) and definitions (i.e. writing new statements). Below, we give explain how individuals, concepts and relations can be defined and used for source code documentation in ACE.

%As we will show in the next section, this is essential to enable high level questions.
%The model has to be manually extended. Part of the extension model are not directly tied to the domain (such as the concepts) and the remaining is Mondrian-specific.

%\paragraph{Individuals}
\emph{Individuals} correspond to logical constants and are the simplest form of ontological entities. They are typically represented in (controlled) natural language as proper names, for example \ace{EventManager}, \ace{Simon}, the \ace{RMoD} group, or the \ace{EventManager Tutorial}:
\begin{aceenv}
EventManager is a system.\\
Simon is a developer.\\
RMoD is a group.\\
The EventManager Tutorial is a user manual.\\
\end{aceenv}
Classes and methods, as they are extracted from the source code, are also individuals and should therefore be represented as proper names.

%\paragraph{Concepts}
Nouns are typically used to represent \emph{concepts}, which correspond to unary predicates in logic. \ace{Class}, \ace{document} and \ace{developer} are examples of such concepts.
In order to have a properly defined knowledge base, the respective concepts should be organized in a hierarchy, called taxonomy. A clean taxonomy helps users and reasoners to interpret and use the given concepts in a coherent way.
Taxonomies consist of subconcept relationships that can be expressed by sentences of the form \ace{every \emph{A} is a \emph{B}}.
%The taxonomy is the backbone of every ontology and should therefore be defined carefully.
This is a simple example of such a taxonomy:
\begin{aceenv}
Every person is an entity.\\
~~~~Every developer is a person.\\
~~~~Every tester is a person.\\
Every group is an entity.\\
Every artifact is an entity.\\
~~~~Every document is an artifact.\\
~~~~~~~~Every user manual is a document.\\
~~~~Every code element is an artifact.\\
\end{aceenv}
Indentation is used here to clarify the hierarchical structure. Typically, some --- but not all --- of the concepts on the same level are supposed to be disjoint. This can be defined with statements such as the following:
\begin{aceenv}
No person is a group.\\
No person is an artifact.\\
No group is an artifact.\\
\end{aceenv}
Of course, this is not the only sensible way to define a taxonomy for source code documentation, and our approach does not depend on a particular one.

%\paragraph{Relations}

There are many ways in which \emph{relations} between individuals, called binary predicates in logic, can be represented in (controlled) natural language, including transitive verbs (e.g.\ \ace{maintain}, \ace{write}), {\em of}-constructs (e.g.\ \ace{member of}), and adjectives with prepositions (e.g.\ \ace{related to}). These are some examples of how such relations can be used:
\begin{aceenv}
If X maintains something then X is a person.\\
If something is a member of X then X is a group.\\
Simon Denier is a member of RMoD.\\
Simon Denier maintains EmergencyHandler.\\
AlarmDesk is written by Simon Denier.\\
The EventManager Tutorial describes EventManager.\\
\end{aceenv}

The general nature of logic allows users to write formal documentation about everything they consider relevant, from modules and bugs over developers and groups to tasks and business plans.

%\begin{aceenv}
%If X maintains something then X is a person.\\
%If something is a member of X then X is a group.\\
%Alexandre Bergel is a member of Pleiad.\\
%The Mondrian Tutorial describes Mondrian and is written by Alexandre Bergel.\\
%If X is related to Y then Y is related to X.\\
%\end{aceenv}

%:==============

\subsection{Queries}\seclabel{queries}

Putting together these different kinds of statements and feeding them all to a logic reasoner, we can automatically answer different kinds of sophisticated questions.
%Asking questions is a natural way of deducing and extracting information from the knowledge base.
Such questions can be written in the same controlled natural language as the other statements. As a start, we can query the source code structure and obtain the answers from the reasoner:
% empty lines removed because it all should belong to the same paragraph.
\begin{aceenv}
Which methods are invoked by more than 80 methods?\\
- Broadcaster-sendMessage\\
- EventMessage-getText\\
- ...\\
\end{aceenv}
\begin{aceenv}
Which classes instantiate a subclass of Event?\\
- AlarmDesk\\
- TemperatureSensor\\
- ...\\
\end{aceenv}

While there exist integrated development environments and other code analysis tools that are able to answer such questions, controlled natural language allows us to formulate them in a more intuitive way. More importantly, users can query the documentation in the same way they can query the code structure, as the following examples show:
\begin{aceenv}
Which members of RMoD maintain more than 20 classes?\\
- Simon Denier\\
- ...\\
\end{aceenv}
\begin{aceenv}
Which classes are related to a class that is maintained by Simon Denier?\\
- AlarmDesk\\
- ...\\
\end{aceenv}
\begin{aceenv}
Which document that describes EventManager is written by a member of RMoD?\\
- the EventManager Tutorial\\
- ...\\
\end{aceenv}
More examples of such questions follow in the subsequent sections.

%:==============

\subsection{Consistency}\seclabel{consistency}

Each new statement is checked by the reasoner --- immediately after its creation --- for consistency. If a certain statement is not consistent with the existing knowledge base, the user is informed and the statement is not included in the knowledge base. In this way, we can be sure that the knowledge base is always consistent, which is a prerequisite for any other reasoning task.

For example, let us consider a situation in which --- among others --- the following statements are present in our knowledge base:
\begin{aceenv}
If X is a direct subclass of Y then X is a subclass of Y.\\
If X is a direct subclass of something that is a subclass of Y then X is a subclass of Y.\\
EventHandler is a direct subclass of Handler.\\
EmergencyHandler is a direct subclass of EventHandler.\\
No subclass of Handler is maintained by a member of Group-A.\\
Every member of Group-B is a member of Group-A.\\
Brian is a member of Group-B.\\
\end{aceenv}
The first two sentences are part of the source code model. The next two are automatically generated from the source code. The last three sentences are part of the user-defined documentation. In this situation, if a user intends to add the statement
\begin{aceenv}
EmergencyHandler is maintained by Brian.
\end{aceenv}
then the reasoner is able to detect that it would introduce inconsistency and can inform the user that this statement cannot be added to the knowledge base. Deliberately, a non-trivial example has been chosen here that is well within the capabilities of state-of-the-art reasoners.

Source code changes are treated the same way: For every change in the source code or its documentation, it is checked whether this change would make the knowledge base inconsistent. If so, the user is informed and has to resolve the issue (by either modifying the source code or the documentation).
   
%:==============
%\subsection{Evolution}\seclabel{evolution}
%
%\ab{Reviewer: a description of what happens when the software changes, and
%      when the documentation changes - is it easy to keep them in
%      sync (there's implications that some of ACE is generated, 
%      but questions as to how much; if it is all, then one would
%      question the usefulness, if it is just the basic facts, then
%      the authors need to talk about evolution of the code base).}
%
%\tk{This could be moved to the Discussion section. We can say that with our approach the changes on the documentation can (at least
%partically) be predicted from intended changes on the source code. E.g. deleting a class would mean deleting several documentation
%statements, which could be shown to the developer as a warning.}
%
%\ab{Can we have a screenshot of an inconsistency?}
%
%\tk{We could. We already give an example of an inconsistency above. Do you think that a screenshot would give additional value?}

%:%%%%%%%%%%%%%%%%%%%%%%%%%%%%%%%%%%%%%%%%%%%%%%%%%%

\section{Implementation} \seclabel{implementation}

We implemented our approach based on AceWiki \cite{kuhn2008swui,kuhn2009semwiki}, an existing semantic wiki engine, in which developers can comfortably browse and edit software documentation written in the ACE language. The writing and improving of the documentation can be done in a collaborative way. AceWiki seems to be well-suited for the evaluation of the general feasibility and usefulness of our approach, but for the future we plan to implement it as an IDE plugin.

The goal of AceWiki is to provide a very intuitive and natural ontology interface, where the complex logic formalisms are hidden behind controlled English. In this way, AceWiki does not only provide a more natural interface, but also supports a higher degree of expressiveness than most existing semantic wikis. The main difference from other semantic wikis --- like Semantic MediaWiki \cite{kroetzsch2007websem} or KiWi \cite{schaffert2009semwiki} --- is the use of controlled English. Articles in AceWiki are written in ACE, a formal language that looks like natural English, but the wiki content is translated into logic in the background, which allows for automatic reasoning. Recently, AceWiki has been extended to support multilinguality \cite{eswc2013_kaljurand}.

Figure \ref{fig:acewikiscreenshot} shows a screenshot of the AceWiki interface. It resembles existing wiki interfaces like the one of Wikipedia. The essential difference is that the content of the wiki articles is not free text, but has to comply with the restrictions of the ACE language.
\begin{figure}[tb]
\begin{center}
\includegraphics[width=\columnwidth]{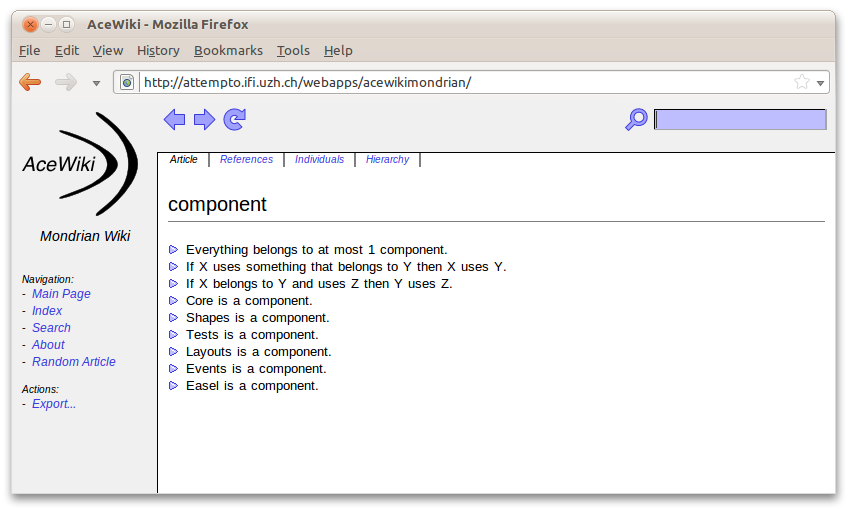}
\caption{A screenshot of AceWiki} \figlabel{acewikiscreenshot}
\end{center}
\end{figure}

In the background, the wiki articles are translated into OWL (Web Ontology Language) \cite{bock2009owl} and given to a reasoner. The reasoner is used to ensure the consistency of the knowledge base and to answer questions. For the case study to be presented, the OWL reasoner FaCT++ \cite{tsarkov2006ijcar} has been used, but others can be used within AceWiki.

%:\ab{I feel this is not a problem, but a particular way of formulating input. I would not be that negative} A major problem of controlled natural languages like ACE is the difficulty to write sentences that comply with the restrictions of the language. While such languages are very easy to read, it is usually much harder to write them.

To assist users writing sentences in ACE, Ace\-Wiki provides a special editor that is able to predict the possible next words or phrases for partial sentences. In this way, users can create well-formed sentences step by step without the need to learn the restrictions of the language in advance. \figref{acewikieditor} shows a screenshot of this editor.
\begin{figure*}[tb]
\begin{center}
\includegraphics[width=\textwidth]{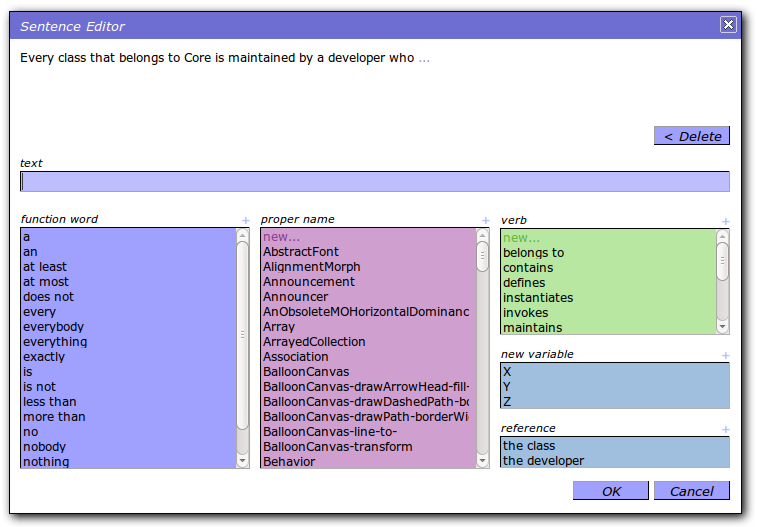}
\caption{The editor of AceWiki} \figlabel{acewikieditor}
\end{center}
\end{figure*}
Apart from function words like \ace{every} or \ace{not} that are predefined and cannot be changed, users can add and edit lexical entries for proper names, nouns, \emph{of}-constructs, transitive verbs, and adjectives with prepositions.

% Even though AceWiki is still under development, it is robust enough to comfortably conduct case studies and user experiments. AceWiki is currently not directly integrated in an IDE, which would be desirable for a future system. At the moment, synchronization is done through file sharing.

%:%%%%%%%%%%%%%%%%%%%%%%%%%%%%%%%%%%%%%%%%%%%%%%%%%%
\section{Case Study} \seclabel{casestudy}

As a case study to evaluate our research question $A$, AceWiki has been employed to document Mondrian,\footnote{\url{http://www.moosetechnology.org/tools/mondrian}} an agile visualization engine written in Pharo,\footnote{\url{http://www.pharo-project.org}} a Smalltalk-inspired programming language. This section introduces Mondrian, describes our documentation effort on it, and shows how this fits into its overall development process.

When Mondrian was released in 2006, its design was clear and neat~\cite{Meye06a}. The growing number of users, however, had triggered an explosion of new requirements that were not originally planned. Satisfying these user requests had been given priority, leaving the documentation behind. After just a few years, the discrepancy between documentation and implementation was considerable. Mondrian's architecture had suffered from the multitude of fixes and feature additions.
\bigskip

\myparagraph{Source Code Model}
We first had to define a source code model for the Smalltalk-style programming constructs of Pharo, which was very straightforward. Based on this model, we extracted from the source code the following four types of statements:
\begin{aceenv}
MORectangleShape is a class.\\
MORectangleShape is a direct subclass of MOShape.\\
MOViewRendererTest-testCachedShape invokes MOShape-isCached.\\
MOViewRendererTest-testCachedShape instantiates MORectangleShape.
\end{aceenv}
%A method is defined using \emph{is a method}, for example \emph{MOShape-isCached is a method}. We use \emph{MOShape-isCached} as the unique identifier to refer to the corresponding method.  A method that belongs to a class is expressed using \emph{defines}, as in \emph{MOShape defines MOShape-isCached}.

Next, we developed an AceWiki exporter for Pharo and the Moose software analysis platform, which we made available under the MIT license.\footnote{\url{http://www.squeaksource.com/AceWikiExporter.html}}
The source code structure is extracted in a simple exchange format and then loaded into AceWiki.\footnote{\url{http://bergel.eu/download/mondrian.acewikidata}} In total, 23\,341 ACE statements were extracted from the Mondrian source code, using the FAMIX metamodel.\footnote{\url{http://www.moosetechnology.org/docs/famix}}
%but any source code information extractor would equally serve (\eg Eclipse's JDT\footnote{\url{http://www.eclipse.org/jdt}}).

% Our ACE translator operates on any object-oriented source code models supported by FAMIX. As a consequence, an application written in Java or C\# project may be represented in ACE. Our approach is also suitable to analyze applications written in a functional programming language by particularizing the translation phase.
\bigskip

\myparagraph{Documenting Mondrian}
Mondrian is a visualization engine that offers a rich domain specific language to define graph-based rendering. Each element of a graph (\ie a node or an edge) has a shape that defines its visual aspects. Nodes may be ordered using a layout. An interactive scripting engine is offered that is called \co{Easel}. The interface with the graphical engine of the host platform is called \co{Morph}.
% Mondrian comes with an extensive set of tests and examples. The examples are accessible from the easel.

Mondrian is an excellent candidate for a documentation effort: We have access to its domain knowledge and its source code; it is a piece of legacy code (older than 6 years); it went through several evolution steps; and it is a critical part of the Moose platform.
Mondrian comes with a reasonable set of documents, including book chapters,\footnote{\url{http://www.themoosebook.org/}} tutorials,\footnote{\url{http://www.moosetechnology.org/tools/mondrian}} research papers~\cite{Meye06a,Lien07b} and mailing list archives. However, these documents are essentially for end users and do not describe the software's internals.
% Some technical discussions were realized via emails, but they lack organization and cannot really be considered as an internal documentation.
%We can safely affirm that before we conducted the work described in this article, the design and architecture of Mondrian was described nowhere. 

%Before guiding the authors of Mondrian to write verifiable documentation about Mondrian's internals,
%we needed a starting point: a reference, even incomplete, that will give the orientation the documentation should take. For that purpose,
%we asked them to roughly identify the main software components and their dependencies.
\begin{figure*}[tbp]
\begin{center}
\includegraphics[width=1.0\textwidth]{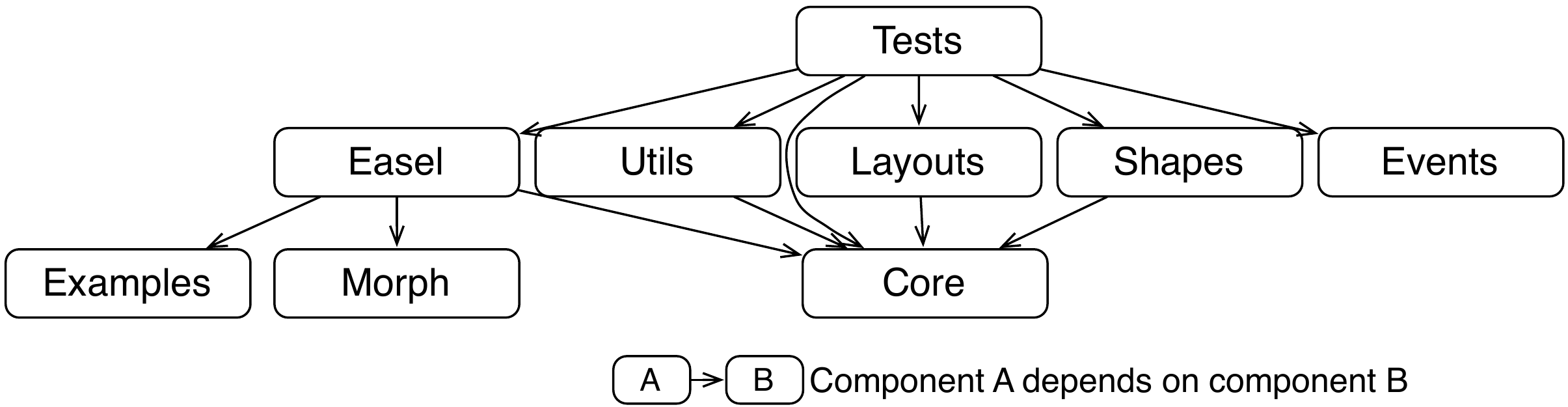}
\caption{Hypothetical architecture of Mondrian} \figlabel{mondrianArchitecture}
\end{center}
\end{figure*}
\figref{mondrianArchitecture} shows the general architecture of Mondrian, as declared by its developers. There are nine components with \co{Core} as the central one.
% We designate a component as a group of cohesive classes, regarded as an elementary unit of the software structure.

%All but \co{Examples} and \co{Morph} depend on \co{Core}.
\bigskip

%:=========
%\subsection{Defining components} \seclabel{definingcomponents}

%\paragraph{Introducing components}
%Mondrian's source code is structured in terms of classes and methods only (\secref{sourcecode}). Components are not represented in Mondrian source code.
\myparagraph{Defining Components}
In order to represent the Mondrian architecture in terms of software components, we first need to define some new words in AceWiki. The terms \ace{component} and \ace{belongs to} were defined to assign classes to components:
\begin{aceenv}
Everything belongs to at most 1 component.\\
If X belongs to Y and uses Z then Y uses Z.\\
\end{aceenv}
Basically, what we do here is to manually extend the model that we introduced above, which contains the concepts \ace{class} and \ace{method} but not \ace{component}. This flexibility to add new concepts and relations allows for expressing things that were not anticipated when the system was designed. This is exactly what source code documentation should mainly do: explain things that cannot be expressed in the given programming language, or only in an implicit way.

We can subsequently define the components that compose Mondrian. First, the individuals need to be defined:
\begin{aceenv}
Core is a component.\\
Shapes is a component.\\
Tests is a component.\\
...
\end{aceenv}
Then, each class can be assigned to a component, using the \ace{belongs to} relation introduced above:
\begin{aceenv}
Every subclass of MOGraphElement belongs to Core.\\
MOGraphElement belongs to Core.\\
MOShape belongs to Core.\\
\end{aceenv}
In this way, class hierarchies and individual classes are assigned to the respective components.
%As a further illustration, The \co{Tests} component is defined as follows:
%\begin{aceenv}
%MondrianTestCase belongs to Tests.\\
%Every subclass of MondrianTestCase belongs to Tests.\\
%\end{aceenv}
%Again, we are exploiting the hierarchies to classify classes since \co{MondrianTestCase} is the root of many unit tests in Mondrian.

%\paragraph{Component dependencies.}
%Now that we have defined components, dependencies may be easily inferred based on the \emph{invokes} relationship (\secref{sourcecode}). We say that a component A \emph{uses} a component B if one method belonging to A invokes a method belonging to B. The \emph{uses} relationship is extended in the following way:
The definition of some relations like \ace{uses} can now be extended to take the new relation \ace{belongs to} into account (and similarly for \ace{class of} and \ace{method of}):
\begin{aceenv}
If X uses something that belongs to Y then X uses Y.\\
\end{aceenv}
%\paragraph{Querying} Queries may be formulated through the AceWiki editor. The definition of \co{Core} is retrieved with the question \ace{What belongs to Core?}.
%The answer is an enumeration of classes that match the facts entered above: \co{MOAbstractLayout}, \co{MOGraphElement}, \co{MO\-Shape}, \co{MOEdge}, \co{MONode}, and few other classes. \co{MOEdge} and \co{MONode}  were not directly mentioned when defining \co{Core}, however these two classes are indirect subclasses of \co{MOGraphElement}. 
%Dependencies between components may be easily asked:
We can now ask different kinds of questions such as
\begin{aceenv}
What belongs to Core?\\
Which component uses Core and uses Examples?
\end{aceenv}
and state architectural properties like:
\begin{aceenv}
No component is used by Core.
\end{aceenv}
This last sentence, however, triggered an inconsistency in our case.
According to \figref{mondrianArchitecture}, this sentence is supposed to be true, and the second question should give \co{Easel} as the only answer. However, as it is often the case when the documentation is not synchronized with the application source code, the exactitude of what is written on a white board is no more than a mere illusion. Asking the question
\begin{aceenv}
Which component is used by Core?
\end{aceenv}
gives us the answers \co{Easel}, \co{Utils}, \co{Layouts}, \co{Shapes} and \co{Events}, which explains why the sentence above triggered an inconsistency. Some dependencies apparently violate the modular design: The \co{Core} component should not depend on other components.

One of the reasons why \co{Core} depends on \co{Layout} can be found in the \co{applyLayout} method defined in \co{MORoot}. \co{MORoot} is a subclass of \co{MOGraphElement}, therefore belonging to \co{Core}. This method is used to order the elements of an object root (it behaves as a composite). Layouts can be applied to an instance of any subclass of \co{MOAbstractLayout}. The source code of \co{applyLayout} is:
\begin{tcode}
MORoot\sepp applyLayout\\
\>"Layout the receiver and all its children."\\
\>self resetMetricCachesResursively.\\
%\>self shapeBoundsAt: self shape put: (0@0 corner: 0@0).\\
\>self do: [ :each \stBar~ each applyLayout ].\\
\>self layout \textbf{applyOn:} self.\\
\>self nodesDo: [:each \stBar\\
\>\>self bounds corner: \\
\>\>\>(self bounds corner\\
\>\>\>\> max: (each extent + each origin)) ]
\end{tcode}
The reason for the odd dependency between \co{Core} and \co{Layouts} stems from the method call indicated in \textbf{bold} in the code shown above.
The method \co{applyLayout} sends the message \co{applyOn:} to the object returned by \co{self layout}, an accessor to the \co{layout} instance variable defined in \co{MOGraphElement}. The method \co{applyOn:} is defined in the class \co{MOAbstractLayout}. An instance of \co{MORoot} therefore directly and explicitly invokes a method defined in the \co{Layouts} component. Whereas having such a method call at runtime is not a problem per se, but explicitly having this call in the source code goes against the stated modularization structure.

This problem was subsequently fixed and included in version 543 of Mondrian. The class \co{MOAbstractLayout} had been moved to the \co{Core} component and renamed as \co{MOLayout}, to follow unique naming patterns.
%(\eg the root of shapes is named \co{MOShape}, and not \co{MOAbstractShape}).
In this new version of Mondrian, \co{Layouts} depends on \co{Core}, but not vice versa.

As another example, the method \co{announce:} of \co{MOGraphElement} is responsible for the dependency of \co{Core} on \co{Events}. This method is used to emit an event (called ``announcement'' in the Pharo terminology), which is dispatched by an announcer. The definition of this method is as follows:
\begin{tcode}
MOGraphElement\sepp announce: anAnnouncement\\
\>"public method"\\
\>announcer ifNil: [ \ret self ].\\
\>anAnnouncement \textbf{element:} self.\\
\>announcer \textbf{announce:} anAnnouncement
\end{tcode}
%Mondrian is written in the Pharo programming language, and Pharo is dynamically typed (\ie variable declaration and method signature do not contains type information).
The message \co{element: self} is sent to the object referenced by the variable \co{anAnnouncement}. The type of \co{anAnnouncement} is not explicit, but the method \co{element:} is defined in the class \co{MOEvent}, which belongs to the component \co{Events}. In addition, the method \co{announce:} is defined in the class \co{MOAnnouncer}, which belongs to \co{Events} too. The dependency of \co{Core} on \co{Events} stems from these two method invocations.

With our approach, improper component dependencies are easily identified by formulating adequate questions:
\begin{aceenv}
Which class of Events is used by a class of Core?\\
Which method of Core uses a class of Events?\\
\end{aceenv}

The classes \co{MOAnnouncer} and \co{MOEvent} are among the results of the first question. \co{MOGraphElement-announce-} is in the answer of the second question. This problem was fixed by moving the method \co{announce:} to the definition of the \co{Events} component. This is consistent with the philosophy of Pharo and Smalltalk to assign methods to different modules rather than their associated classes. This mechanism is commonly designed as ``class extension''~\cite{Berg05a}.

About ten anomalies were identified in Mondrian's architecture in the form of improper component dependencies. Versions 511 to 544 of Mondrian fixed all of them, making its architecture significantly cleaner.\footnote{All the different versions and their associated comments are available online on \url{http://www.squeaksource.com/Mondrian.html}}
\bigskip

%:=========
%\subsection{Document and Source Code Synchronization}\seclabel{synchronization}

\myparagraph{Code and documentation synchronization}
With our approach, developers are notified when some parts of the documentation are invalidated by a certain change in the source code. This is illustrated by the documentation effort on Mondrian that began with version 511. The definition of components was the first action we realized. The \co{Shapes} component, for example, was simply defined as all subclasses of \co{MOShape}:
\begin{aceenv}
Every subclass of MOShape belongs to Shapes.
\end{aceenv}
At that time, the two questions
\begin{aceenv}
Which method of Shapes uses Events?\\
Which method of Shapes uses Layouts?
\end{aceenv}
returned an empty answer, which is what one would expect according to the initial design of Mondrian (\figref{mondrianArchitecture}). With the following two statements, we had forbidden unwanted dependencies:
\begin{aceenv}
No method of Shapes uses Layouts.\\
No method of Shapes uses Events.
\end{aceenv}
However, changes were made over time, and version 525 introduced the method \co{display:on:} to the class \co{MOChildrenShape}:
\begin{tcode}
MOChildrenShape\sepp display: anElement on: aCanvas\\
\>"The element to display can only have\\
\>~~a formshape as a shape"\\
\\
\>\stBar~ builder bounds \stBar\\
\>...\\
\>builder := anElement shape builder.\\
\>bounds := builder \textbf{boundsOf:} self.\\
\>...
\end{tcode}
This method belongs to \co{Shapes} but calls \co{boundsOf:}, a method that returns the geometrical shape and location of graphical elements. \co{boundsOf:} is defined by the class \co{MOFormsBuilder}, which belongs to \co{Layouts}. Therefore, a dependency of \co{Shapes} on \co{Layouts} is introduced, which is inconsistent with the documentation statement \ace{No method of Shapes uses Layouts}. Programmers were notified by the reasoner about this problem, and version 544 of Mondrian is now consistent again.
\bigskip

\myparagraph{What we have gained}
Before our effort, Mondrian lacked documentation about its internal structure and contained a number of unknown structural errors. Now, its architecture and its components are described by entirely formal statements that are at the same time as easy to read and understand as natural English.
%Our documentation is accessible online and may be navigated through from any web browser\footnote{\url{http://attempto.ifi.uzh.ch/webapps/acewikimondrian/}}.
%Mondrian is an essential part of the Moose platform. Its development begun in 2006, more than 8 programmers contributed to its source code since April 2010, and it is intensively used in many independent projects. Its development process is well established and makes an intensive use of agile methods. 
Our documentation effort has been realized without interrupting the development of Mondrian. We made a direct, positive impact on the developers, seeing some engineering tasks and discussions on mailing lists in response to our discoveries.
%(cf. \secref{definingcomponents} and \secref{synchronization}).

%:%%%%%%%%%%%%%%%%%%%%%%%%%%%%%%%%%%%%%%%%%%%%%%%%%%

\section{Experiment} \seclabel{experiment}

The case study presented above indicates that our approach is feasible, but it does not tell us whether controlled natural languages are better or worse than other formal languages. For that reason, we performed a user experiment to find out how controlled natural language compares to other formal languages when used for source code documentation. The goal of this experiment is to assess how easy or difficult it is to understand such documentation statements in different formal languages.

This section investigates research question $B$. In contrast to question $A$, we are \emph{not} using the Mondrian example to investigate this issue, which might require some explanation. For relying on Mondrian in such an experiment, we would have to let participants get familiar with Mondrian or we would have to only select participants that already know Mondrian sufficiently well. In the first case, preparation time would have to be much longer (probably several hours) than the time participants are usually available for. In the second case, it would be very hard to control for the previous experiences of the participants with the given exemplary code base, and it would be difficult to disentangle the effects of our approach with other types of documentation the users might have been in contact with before. Also, it would be virtually impossible to get enough participants with the required expertise. For these reasons, we decided to base the experiment on neutral and ficticious scenarios, presented as schematic diagrams.

Besides ACE, we chose the SOUL syntax (as shown on the official SOUL website\footnote{\url{http://soft.vub.ac.be/SOUL/}}) and Prolog syntax as the languages to be tested against each other. The latter two languages have both been applied to query and document source code structure \cite{wuyts1998tools,kramer1996wcre,Wuyt01a,Fabr03a}, and are the state of the art in this respect. Supplementary material of the experiment is available online.\footnote{\url{http://bit.ly/AceWikiExperimentMaterial}}

\subsection{Experiment Design}

The experiment design is based on a modified version of the ontograph framework \cite{kuhn2009cnlmain}. The basic idea is to use simple diagrams as a neutral basis for comparison. Diagrams have the advantage that they are very different from textual languages, which minimizes the possible bias towards one of the languages to be tested. One of the four diagrams used in the experiment is shown in Figure \ref{fig:testing}. These diagrams show several components of different types and with different relations between them in a very simple graphical notation. They represent a very simple abstract model of a software system, which we can used to evaluate and compare the three languages we have chosen. For each of these diagrams, twenty statements had been written in each of the three languages in a way that each statement has semantically equivalent statements in both other languages. Some of these statements are true with respect to the diagram, others are false. The experiment is designed in a way that each participant is tested on all three languages. The order in which the languages are presented is shuffled in a balanced way.
\begin{figure*}[tbp]
\begin{center}
\includegraphics[width=\textwidth]{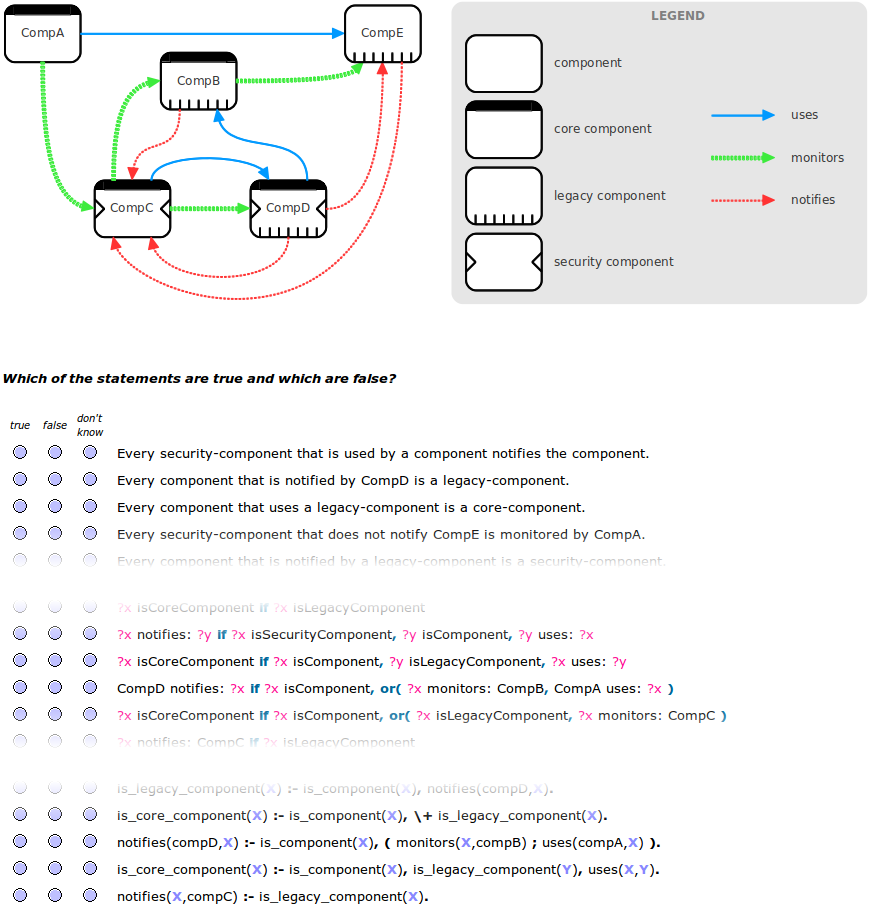}
\caption{This figure is a blending of three screenshots (one for each of the three tested languages) of the testing phases of the experiment. Participants had to classify the statements at the bottom with respect to the diagram shown at the top.}
\figlabel{testing}
\end{center}
\end{figure*}

During a learning phase of at most ten minutes, a short explanation of the language is shown to the participants together with a diagram and five true statements (in the respective language) marked as ``true'' and five false statements marked as ``false''. After the learning phase, there is a testing phase of up to six minutes, during which the participants are presented another diagram and ten statements in the respective language. The objective of the participants is to classify each of the statements as ``true'' or ``false'' (they can also choose ``don't know''). \figref{testing} shows how the screen looks during the testing phase. These learning and testing phases are repeated for each of the languages with different, but similar diagrams for the testing phases (these diagrams for the testing phases can be transformed into each other by a simple reordering and replacement of names, concepts and relation types). Thus, each of the twenty participants has to go through three testing phases consisting of ten statements each, which leads to overall 600 data points in the form of correct or incorrect classifications. After the three learning and testing phases, the participants have to fill out a short questionnaire, asking about their background and their experiences during the experiment.

Twenty experienced graduates and advanced undergraduates in the area of software engineering were recruited as participants for the experiment. They were on average, 27 years old, and all but one of them were male. Most of them (70\%) had never used before a formal language similar to ACE; 60\% and 45\% had never used a language like SOUL or Prolog, respectively.

\subsection{Experiment Results}

The design of the experiment allows us to evaluate the differences of the three languages along three dimensions: the classification score, the time needed, and the preference of the participants derived from the questionnaire. Figure \ref{fig:evaluationresults} shows the results.

\begin{figure}
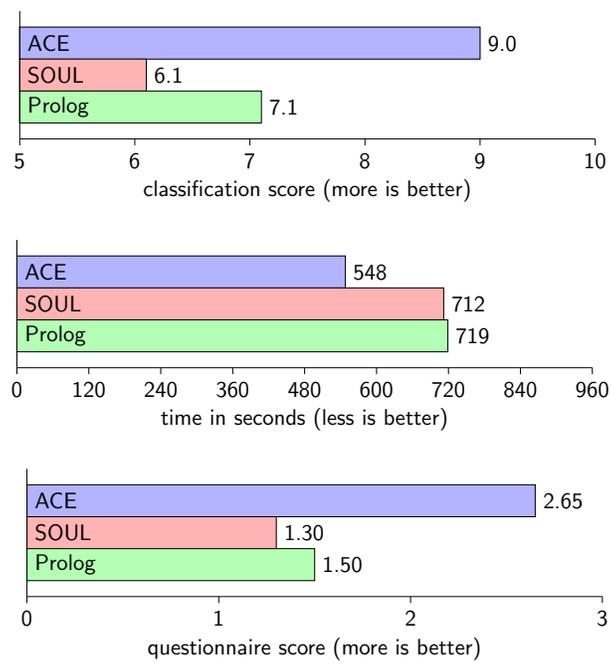

\begin{center}
\scalebox{0.85}{
\begin{bchart}[min=5,step=1,max=10,width=9cm]
  \bcbar[text=ACE,color=blue!30]{9.0}
  \bcbar[text=SOUL,color=red!30]{6.1}
  \bcbar[text=Prolog,color=green!30]{7.1}
  \bcxlabel{classification score (more is better)}
\end{bchart}}
\bigskip\par
\scalebox{0.85}{
\begin{bchart}[step=120,max=960,width=9cm]
  \bcbar[text=ACE,color=blue!30]{548}
  \bcbar[text=SOUL,color=red!30]{712}
  \bcbar[text=Prolog,color=green!30]{719}
  \bcxlabel{time in seconds (less is better)}
\end{bchart}}
\bigskip\par
\scalebox{0.85}{
\begin{bchart}[step=1,max=3,width=9cm]
  \bcbar[text=ACE,color=blue!30]{2.65}
  \bcbar[text=SOUL,color=red!30]{1.30}
  \bcbar[text=Prolog,color=green!30]{1.50}
  \bcxlabel{questionnaire score (more is better)}
\end{bchart}}
\end{center}
\caption{The results of the experiment concerning classification score, time, and questionnaire score for ACE, SOUL and Prolog.}
\figlabel{evaluationresults}
\end{figure}

The most important dimension is the classification score, i.e. the number of correctly classified statements out of ten. Since it is just a binary choice, a score of 5 can be achieved by mere guessing. ``Don't know'' answers and missing answers (when the time limit ran out) are counted as halfway correct (0.5). With ACE, the participants managed to correctly classify an average of 9 out of 10 statements. With SOUL and Prolog, the average score was only 6.1 and 7.1, respectively. The dominance of ACE shows high statistical significance (see below). Thus, the participants understood ACE much better than SOUL or Prolog.

Looking at the time dimension, we see that ACE performed again clearly and significantly better than the other two languages. To learn ACE and to finish the classification task, the participants needed on average about 9 minutes out of 16 (a maximum of 10 for the learning phase plus 6 for the testing phase). In contrast, they needed about 12 minutes in the case of SOUL and Prolog.

The third dimension is derived from the questionnaire, which contains a question ``How easy or hard to understand did you find the statements of language X?'' for each of the languages. The possible answers are ``very easy to understand'' (value 3), ``easy to understand'' (2), ``hard to understand'' (1), and ``very hard to understand'' (0). ACE was perceived on average as being between ``easy'' and ``very easy'' to understand with a tendency to the latter. Prolog, on the other hand, was in the middle of ``hard'' and ``easy'', and SOUL even a bit lower. Thus, ACE was perceived as much more understandable than the other languages.

To check the observed differences between the languages for statistical significance, we use the Wilcoxon signed rank test. This is a non-parametric test to check whether or not we can conclude from observed differences within a paired sample that the difference exists in reality. This test method is robust in the sense that it does not require the statistical population to be normally distributed. Table \ref{tab:significance} shows the obtained $p$-values. These $p$-values represent the probability that the observed differences result by chance from a population where no differences are present. The lower these values, the more confident we can be that a difference exists in reality.
%\ab{What is the maximum p can be? 1? How much should be expect for a random selection?}
All differences involving ACE are statistically significant on a 95\% confidence level (and would even be significant on a 99.9\% level). The differences between SOUL and Prolog are not significant.

\begin{table}
\caption{The $p$-values for the differences between the three languages, using Wilcoxon signed rank tests.}
\begin{center}
\begin{tabular}{r|l|l|l}
& ACE$\sim$SOUL & ACE$\sim$Prolog & SOUL$\sim$Prolog \\
\hline
classification & 0.0000076 & 0.000092 & 0.091 \\
time & 0.00012 & 0.000038 & 0.62 \\
questionnaire & 0.00011 & 0.00069 & 0.30 \\
\end{tabular}
\end{center}
\tablabel{significance}
\end{table}

In addition to the three dimensions explained above, the questionnaire contains the question ``If you were a software developer who has to document a piece of software, which formal language would you prefer to use?''. 65\% of the participants chose ACE, 20\% chose SOUL and 15\% chose Prolog. Thus, they did not only consider ACE to be more understandable, most of them would also prefer to use ACE for \emph{writing} source code documentation.

In summary, ACE was understood much better, required less time to learn, and was preferred by the participants. In all explored dimensions, ACE clearly outperformed the other languages.

%:%%%%%%%%%%%%%%%%%%%%%%%%%%%%%%%%%%%%%%%%%%%%%%%%%%
\section{Discussion} \seclabel{discussion}

%=========
\subsection{Applicability to Other Programming Languages}

Our results are based on the experiments we have carried out using the Pharo programming language, but our general approach is not restricted to a particular language.
%\paragraph{Source code modeling} 
%Modeling in ACE
In our source code model, we have considered the following language elements: packages, classes, methods, invocations between methods and class instantiation. These elements are sufficient to capture the general architecture and dependencies of Pharo programs, but they also apply to many other (object-oriented) programming languages.

It is relatively easy to adopt the approach to other programming languages (but this would typically be a task of the tool developer and not of the end user). For example, to add the concept of \emph{interfaces}, as e.g.\ Java supports it, the source code model could be extended by the following statements:
\begin{aceenv}
Every interface is a code element.\\
No interface is a class.\\
No interface is a method.\\
If X implements Y then X is a class.\\
If X implements Y then Y is an interface.\\
\end{aceenv}

%Parsing
From this, the ACE parser can automatically generate the formal model, which can be mapped to the source code structure as provided by tools such as the Moose software analysis platform\footnote{\url{http://moosetechnology.org}}, which offers a wide range of language parsers, including Pharo, VisualWorks, Java, C, Cobol, Delphi, and C++.
%Modeling using Moose and FAMIX
Additionally, we use the FAMIX meta-model, which provides a language-independent model of programming language elements~\cite{Deme01y}. It includes concepts such as a class, method, package, and invocation, and is general enough to cover most structural and behavioral elements commonly found in programming languages.

%Generating facts
Once these mappings are in place for a particular language such as Java, we can represent source code in ACE:
\begin{aceenv}
Runnable is an interface.\\
Thread is a class.\\
Thread implements Runnable.
\end{aceenv}
These facts introduce the elements necessary to later attach documentation statements and to provide a mapping to the given FAMIX model.  

%\paragraph{Fact definition}
%The approach proposed in this paper involves a reification of the application source (\ie transforming source code contained in text files into elements processable by the logic reasoner) and the interpretation of facts written in the ACE controlled natural language. 

%=========
\subsection{Keeping queries simple}

During our documentation effort, we had to query the source code of Mondrian multiple times in order to pinpoint anomalies. Our experience shows that we had to face a number of complex queries, making heavy use of subordinate clauses introduced with the conjunction \ace{that}.

For example, to identify the causes for the dependencies between \co{Core}, \co{Layout} and \co{Events} we first formulated these questions:
\begin{aceenv}
Which class that belongs to Layout is used by a class that belongs to Core?\\
Which method that is defined by a class that belongs to Core uses a class that belongs to Layout?\\
Which method that is defined by a class that belongs to Core uses MOPortEvent?
\end{aceenv}
Although these sentences are grammatically correct from a linguistic point of view, they are difficult to read. To reduce the complexity of these questions, we defined the relations \ace{method of} and \ace{class of}, which have already been mentioned above. In the case of \ace{method of}, we first had the following facts to express the relation between a method and its class:
\begin{aceenv}
If X is a method of Y then X is a method.\\
If X defines Y then Y is a method of X.
\end{aceenv}
We later extended the relation to cover the \ace{belongs to} relation:
\begin{aceenv}
If X is defined by something that belongs to Y then X is a method of Y.
\end{aceenv}
Thus, apart from connecting methods and classes, the relation \ace{method of} can now also relate methods and components. This simple statement greatly simplifies the formulation of method belonging. The relation \ace{class of} has been defined in a similar way. The three questions above can now be formulated in a much simpler way:
\begin{aceenv}
Which class of Layout uses a class of Core?\\
Which method of Core uses a class of Layout?\\
Which method of Core uses MOPortEvent?
\end{aceenv}

\subsection{Kind of documentation}

%We demonstrated that a documentation about software structure and design can be efficiently written in ACE and processed with AceWiki. However, we have focused on documentation that is closely related to the software source code. So far, we haven't intended writing in ACE a different kind of documentation, like a tutorial or manual for end-user.

In order to illustrate that various kinds of documentation statements can be expressed with our approach, we take the ten design entity attributes as defined by the IEEE standard 1016-1998 \cite{ieee1016-1998}: Identification, Type, Purpose, Function, Subordinates, Dependencies, Interface, Resources, Processing and Data. Below, we exemplify how each of them can be expressed, at least partially, in ACE:

\begin{itemize}

\item Unique identification is done by assigning a name, \eg ``Pleiad group'', to an entity by adding it to the AceWiki lexicon.

\item Types like subsystem, module, process, algorithm or data store are defined as nouns in the AceWiki lexicon and entities can be assigned to these types by simple ``\ace{is a}''-sentences:
\begin{aceenv}
Core is a module.\\
The Medical Database is a data store.\\
\end{aceenv}

\item Even though the purpose of a certain entity cannot be fully defined in ACE (it is a complex issue to formally represent purposes), it is very easy to give names to certain purpose aspects and to link them to the entities:
\begin{aceenv}
A purpose of ResultsCache is performance improvement.\\
A purpose of ResultsCache is system stability.\\
\end{aceenv}

\item The function of an entity like input, transformations, and output can be documented:
\begin{aceenv}
The Attempto Parsing Engine processes Attempto Controlled English and returns the ACE DRS format.\\
The Medical Database stores every patient record.\\
\end{aceenv}

\item Subordinate entities can be expressed in AceWiki with relations like \ace{contains} or \ace{part of}:
\begin{aceenv}
MySystem contains exactly 10 modules.\\
Every module is a part of MySystem.\\
Core contains the Data Base Manager.\\
\end{aceenv}

\item Dependencies can be expressed as follows:
\begin{aceenv}
ModuleA depends on ModuleB.\\
Every subclass of QueryWindow depends on a subclass of DataBase.\\
\end{aceenv}

\item Much information about interfaces, \eg which methods to call from the outside and which parameters to use, is already present in the source code in many cases. However, AceWiki allows for additional documentation:
\begin{aceenv}
Every method that invokes DataStore-aquireLock invokes DataStore-releaseLock.\\
Every method that invokes a test method belongs-to the Test module.
\end{aceenv}

\item Dependencies on external resources can be defined as follows:
\begin{aceenv}
The Data Base Manager requires MySQL.\\
The Receipt Generator requires a printer.\\
\end{aceenv}

\item The processing attribute is about how an entity achieves its function, \eg which algorithms it uses:
\begin{aceenv}
ListSorter uses the Quicksort algorithm.\\
\end{aceenv}

\item The data attribute should contain information about the internal data structures:
\begin{aceenv}
ElementDirectory uses a hash table.\\
\end{aceenv}

\end{itemize}
Thus, ACE and AceWiki allow for the documentation of all ten of these design entity attributes, at least in principle and to a certain degree. In some cases like for the purpose attribute, it is not possible to give a full description in ACE, because it is difficult to represent such things in logic. But, as the examples show, even in these cases, it is possible to name the involved concepts and relate them to the affected individuals.

%=========
\subsection{Performance} \seclabel{performance}

To measure the performance of AceWiki for larger code bases, we micro-benchmarked our approach on seven applications, including ArgoUML,\footnote{\url{http://argouml.tigris.org}} a popular large software for case studies. The amount of facts to describe these applications ranges from around 1\,400 to more than 170\,000 (ArgoUML). Mondrian leads to around 23\,000 facts.

The results are presented in \figref{benchmarks}. For each of the applications, it shows the time needed (i) to load and parse the source code, (ii) to add a fact to the knowledge database, (iii) to check the consistency of the knowledge database, and (iv) to detect an inconsistency.

\begin{figure*}[tbp]
\begin{center}\footnotesize\scalebox{0.98}{
\begin{tabular}{@{~}r@{~}|@{~}r@{~}|@{~}r@{~}|@{~}r@{~}|@{~}r@{~}|@{~}r@{~}|@{~}r@{~}|@{~}r@{~}|}
\textbf{Software}	& \textbf{\#classes}	& \textbf{\#methods}	& \textbf{\#facts}	& \textbf{load facts}	& \textbf{add fact}	& \textbf{check for} & \textbf{detect}\\
  & & & & & & \textbf{consist.} & \textbf{inconsist.}\\
\hline
Adore & 12 & 163 & 1\,442 & 14.12 & 0.019 & 0.02 & 0.07 \\
Gofer & 32 & 234 & 2\,276 & 20.05 & 0.15 & 0.46 & 0.06 \\
Autotest & 56 & 200 & 1\,701 & 16.22 & 0.014 & 0.09 & 0.24 \\
Mondrian & 148 & 1\,902 & 23\,342 & 25.56 & 0.55 & 0.55 & 4.88 \\
Glamour & 157 & 1\,286 & 17\,798 & 159.35 & 0.94 & 0.44 & 5.26 \\
Moose & 357 & 4\,218 & 75\,021 & 2774.17 & 1.98 & 0.76 & 6.40 \\
ArgoUML & 2\,526 & 20\,338 & 173\,982 & 7835.25 & 2.40 & 2.50 & 14.76 \\
\hline
\end{tabular}}
\end{center}
\caption{Micro benchmarks of AceWiki (the unit of time for the last four columns is seconds)}\figlabel{benchmarks}
\end{figure*}

The figures show that the OWL reasoner FaCT++ behaves relatively well when adding facts and checking for consistency. Adding a fact to a database with 1\,442 facts takes 0.019 seconds. For the largest application, it takes 2.40 seconds.
Loading the knowledge base is much slower. It takes about 14 seconds for 1\,441 facts, 25 seconds for Mondrian and more than 7\,800 seconds for ArgoUML. This is because the reasoner requires the complete ontology to be loaded into memory in order to perform reasoning tasks. However, there is certainly a lot of room for improvement in this respect, \eg by adopting techniques used by highly scalable triple stores or by excluding some of the most complex OWL features, which typically lead to a drastic drop in time consumption.

%Loading the database is a cost that is incurred only once, when the source code is first entered in the reasoner. Even if it takes nearly 1 hour to load a very large database, this is a cost that appear to us as reasonable since in practice it is incurred just once.
%Subsequent source code modification are captured as a set of fact addition, which a much shorter execution time.

%\ab{from the reviewer:     - a discussion of the cost of analysis of all the facts;
%      whether it is wholesale or incremental; 7800 seconds
%      sure sounds like a lot - what are the implications?}\ab{I worked at the text that mitigate this.}
%\tk{I changed it a little. It is hard to argue that 7800 seconds is not a lot. So, I changed the line of argument.}

AceWiki did not encounter problems loading a larger knowledge base consisting of more than 23\,000 statements. We were not able to precisely measure the memory consumption, but we experienced a consumption of 817 MB of memory to construct and load the Mondrian database. Considering the resources available in modern computers, this does not seem to be a critical issue. This benchmark was carried out on an Ubuntu Linux 2.6.32-24 with 2GB of RAM and a 2.4 GHz Intel Core 2 Duo.

%\subsection{Documentation vs. annotation}
%
%A very important property of logic languages like OWL and ACE is the fact that they are domain-independent. The laws of logic are rules that apply to any domain. For this reason, AceWiki users are in general not restricted on a certain domain but can write statements about everything. An AceWiki instance for source code documentation may contain statements that describe many concepts, including  versions, developers, groups, change proposals, problems, bugs. In this sense, AceWiki allows not only for \emph{annotation} of the source code elements but for a proper \emph{documentation} of the concepts behind the source code and its environment.
%
%The only restriction is the expressiveness of OWL and ACE. But even if this expressiveness is not sufficient to fully express a certain definition or state of affairs, it is still possible to name the respective entities and to define some simple relations between them.

%=========
\subsection{Usability}

The Mondrian case study described above indicates that AceWiki is suitable for the representation of software documentation. The results of the experiment show that ACE statements are easy to understand for software engineers and that they prefer ACE over other formal languages to document source code. One question that is not answered by the presented studies, however, is whether developers are \emph{able to write} sensible ACE statements with AceWiki. This aspect has been researched in previous work.

Two usability experiments for AceWiki have been performed with altogether 26 participants \cite{kuhn2009semwiki}. The results showed that AceWiki and its editor are easy to learn and use. The participants --- without being trained to use AceWiki or ACE --- managed to add many statements to AceWiki in a short period of time. On average, each user created a correct and sensible statement every 5--6 minutes, \emph{while} defining a new word every 5--7 minutes. 78\%--81\% of the statements were correct and sensible, and 61\%--70\% of them were complex in the sense that they contained negations, implications, disjunctions, or number restrictions. The participants were mostly students, with no particular background in software engineering. Since software engineers are used to working with complex software tools, it is reasonable to assume that they perform at least as well as average students. Another study confirmed that writing ACE sentences with the editor used in AceWiki is easier and faster than writing in other formal languages \cite{kuhnhoefler2012coral}.

Another experiment with 64 participants evaluated the understandability of ACE, independently of a particular tool \cite{kuhn2013swj}. The results showed that ACE is not only easier to understand than a comparable logic language but also needs less time to be learned and is preferred by users. Under a strict time limit, the participants managed to correctly classify more than 90\% of given ACE statements (which semantically covered a broad subset of OWL) as true or false with respect to certain situations. The participants were students or graduates with no higher education in computer science or logic. Again, we believe that software developers can be expected to be at least as skilled in logical thinking as average students, and should therefore be able to correctly interpret ACE statements with a high accuracy.

%\ab{We can address these issues in a journal version maybe}
%Plurial and ...of
%What is a direct subclass of MOGraphElement  vs What are the direct subclasses of MOGraphElement

%Open world assumption
%
%Cardinality and chaining. For example, we cannot use cardinality of uses since uses is a deduction of existing relations. This is a limitation of OWL. 
%
%We cannot prove negation. "Which classes do not depend on Core?"
%
%AceWiki does not support adjectives

%=========

%\subsection{Applicability to other languages}
%
%Mondrian, the case study presented in this article, is written in Pharo. Pharo is a dynamically typed languages, implying that variables and method signatures are defined or declared without any type annotation. The analyzes realized on Mondrian with AceWiki are therefore shaped by the characteristics of Pharo. 
%
%The queries formalized in ACE operate on a Moose model. Queries are operated on Moose entities and involve features restricted to Pharo. A Moose model is composed of Famix entities. Famix is a language-independent metamodel, therefore able to describe programs written in a statically-typed language. 
%
%ACE accommodates to the use of type related queries. Consider the following ACE facts to model the type relation in Java:
%
%\begin{aceenv}
%Object is a type.
%Every class has a type.
%...
%\end{aceenv}
%
%\ab{How can we say that a class may implement one or more interfaces?}
%
%Our approach is applicable to programming languages other than Pharo since relation between entities are easily 

%:%%%%%%%%%%%%%%%%%%%%%%%%%%%%%%%%%%%%%%%%%%%%%%%%%%
\section{Related Work} \seclabel{relatedwork}

The use of logic to describe and check software architecture has been explored in depth \cite{mens1999tools,wuyts2001phd}. In this context, a number of different tools have been implemented, \eg Reflexion Models \cite{murphy1995sigsoft}, ArchJava \cite{aldrich2002icse}, Lattix Inc's dependency manager \cite{sangal2005oopsla}, and Intentional Views \cite{mens2006clss}. However, in contrast to the approach presented here, the resulting formal models of the architecture cannot be read and queried in a natural way and, as far as we are aware of, they have not been used in practice to \emph{document} software. Users are supposed to learn how to read and write statements in some sort of formal logic. This could be a major hindrance for the broad adoption of such systems, especially if the model of the architecture has to be verified by non-computer-scientists.

The use of wikis for software documentation is quite common. There even exists a wiki engine --- Dokuwiki --- that is specifically dedicated to documentation \cite{badger2006eosm}. However, the documentation cannot be checked for consistency or be used to answer specific questions since such wikis lack semantic capabilities.

W\"ursch et al. \cite{wuersch2010icse} present an approach to use controlled natural language in the context of software engineering that is in some respects very similar to our approach. Their system allows developers to ask questions about their source code in a controlled language. The difference to our approach is that only questions are supported. There is no possibility to augment the underlying model by annotations or documentation statements.

Kaplanski \cite{kaplanski2010cnl} also explored the use of controlled English to model object-oriented systems, which is very similar to our approach in many respects. However, he does not provide evaluation results or a running prototype at this point.

%\ab{Reviewer: The related work is missing the major work in automatically documenting   code. }\ab{I added some more}
  
Kimmig \etal\cite{Kimm11a} propose an approach for querying source code using a natural language. Users write queries in a simple question that follows the pattern question word - verb - noun - verb (\eg ``Where is balance read?'' with ``balance'' being an instance variable). Our approach differs on two points: (i) With our approach, new concepts and relations are easy to define. Their approach is fairly limited in that respect. (ii) They apply a sophisticated procedure of cleaning and tokenizing the natural language queries, before formalizing them. With our approach, documentation is written in a precise and unambiguous language in the first place.
 
Buse and Weimer~\cite{Buse10a} synthesize succinct human-readable documentation from software modifications. They employ an approach based on code summarization and symbolic execution, summarizing the runtime conditions necessary for the control flow to reach a modified statement. It is unclear, however, whether such techniques are useful for more than helping to write version log messages.

%:%%%%%%%%%%%%%%%%%%%%%%%%%%%%%%%%%%%%%%%%%%%%%%%%%%
\section{Conclusion and Future Work} \seclabel{conclusion}

To conclude, we can come back to the three challenges we identified in \secref{motivbackgr} for a documentation system based on ACE to be effective and reliable, and we can review these challenges against our approach:
\bigskip

\noindent\textbf{Validation.}
With our approach, the consistency of the documentation can be automatically checked by an OWL reasoner. The documentation is incrementally validated when new facts are added or removed.
%This can be done at the same pace new documentation and new source code are written.
\bigskip

\noindent\textbf{Immediate feedback.} 
After the initial loading of the whole code source and documentation data, our approach considers only incremental modifications. As we have shown, the cost of adding a new fact is only a few seconds or less. We can reasonably assume that the code production rate of a software developer remains below the processing time of our approach. Our system is therefore able to quickly provide feedback for any given modification.
%In case of a multi-developers setting, the documentation validation process could be synchronized with a source code manager.
\bigskip

\noindent\textbf{Minimal impact on software engineers.} 
In our approach, the documentation is written in ACE, a language that can be learned quickly. Our experiment showed that ACE is easy to understand and preferred by developers over other languages. Ideally, the documentation statements could be embedded in the source code as comments and supported by the IDE. This minimizes the impact on software engineers.
\bigskip

The importance of writing documentation about how a software operates does not need to be demonstrated. However, this activity is rarely a priority when not made an explicit requirement. In this article we showed that a reliable solution may be achieved with a controlled natural language interpreter and facilities for checking consistency and answering questions. We showed that controlled natural language is easier to understand than other languages, needs less time to be learned, and is preferred by the users.
Our approach validates documentation against the actual source code, provides immediate feedback, and leads to only a small amount of additional work for software engineers.

Agile methods promote the use of unit tests as an executable kind of documentation. In the same spirit, we consider our work as a small step towards the vision of a fully automated code documentation system.

For the future, we plan to propose a concrete format to include ACE statements in source code comments, and to provide an IDE plugin. The relevant features of AceWiki should be included in this plugin and a reasoner should be integrated. In this way, the IDE can help developers to write ACE sentences, notify them about problems after changes in the source code or the documentation, and answer questions about the code structure and its documented environment. In addition, we plan to include an explanation module, which some OWL reasoners provide, in order to show to the developers \emph{why} a particular question gave the respective answers or \emph{why} a statement triggered inconsistency. This system should be designed in a way that hides the complexity of formal logic and feels like a natural and intuitive extension of the programming environment. We plan to evaluate the impact of such documentation tools on open source communities, particularly in the agile programming field.

% Asking questions about a system
%\paragraph{Program understanding} Gail Murphy in one of her papers said that most developers have 35 questions when they develop . Novice developers 

\section*{Acknowledgment}

We would like to thank Tudor Doru G\^irba, Roel Wuyts and the whole Pleiad research group for their feedback on an early draft of this article. We also thank all the people who participated in our experiment.

We gratefully thank Renato Cerro for revising an earlier version of this article.

This work has been partially funded by Program U-INICIA 11/06 VID 2011, grant U\
-INICIA 11/06, University of Chile, and FONDECYT project 1120094.

\bibliographystyle{elsarticle-num}
\bibliography{main,scg}

\end{document}